\documentclass[conference]{IEEEconf}

\input epsf
\usepackage{url}
\usepackage{xspace}
\usepackage{hyperref}
\usepackage{todonotes}
\usepackage{color,xcolor}
\usepackage{amsmath,amssymb}
\usepackage[most]{tcolorbox}

\newcommand{\eg}{\textit{e.g.,}\xspace}
\newcommand{\etc}{\textit{etc.}\xspace}
\newcommand{\etal}{\textit{et al.}\xspace}

\newcommand{\secref}[1]{Section~\ref{#1}\xspace}

\newcommand{\gpt}{{ChatGPT}\xspace}

\newtcolorbox{GPT}[2][]{colback=gray!10!white, colframe=black, fonttitle=\bfseries, colbacktitle=gray!100, enhanced,breakable, attach boxed title to top center={yshift=-2mm}, title={$\blacktriangleright$~#2~$\blacktriangleleft$}, #1}

\renewcommand\thesection{\arabic{section}}
\renewcommand\thesubsectiondis{\thesection.\arabic{subsection}.}
\renewcommand\thesubsubsectiondis{\thesubsectiondis.\arabic{subsubsection}.}

\begin{document}

\title{LLM for Test Script Generation and Migration: \\ Challenges, Capabilities, and Opportunities}

\author{Shengcheng Yu, Chunrong Fang$^{*}$, Yuchen Ling, Chentian Wu, Zhenyu Chen$^{*}$\\
	\normalsize State Key Laboratory for Novel Software Technology, Nanjing University, Nanjing, China\\
	\normalsize *Corresponding authors: \{fangchunrong, zychen\}@nju.edu.cn
}

\maketitle

\begin{abstract}

This paper investigates the application of large language models (LLM) in the domain of mobile application test script generation. Test script generation is a vital component of software testing, enabling efficient and reliable automation of repetitive test tasks. However, existing generation approaches often encounter limitations, such as difficulties in accurately capturing and reproducing test scripts across diverse devices, platforms, and applications. These challenges arise due to differences in screen sizes, input modalities, platform behaviors, API inconsistencies, and application architectures. Overcoming these limitations is crucial for achieving robust and comprehensive test automation.

By leveraging the capabilities of LLMs, we aim to address these challenges and explore its potential as a versatile tool for test automation. We investigate how well LLMs can adapt to diverse devices and systems while accurately capturing and generating test scripts. Additionally, we evaluate its cross-platform generation capabilities by assessing its ability to handle operating system variations and platform-specific behaviors. Furthermore, we explore the application of LLMs in cross-app migration, where it generates test scripts across different applications and software environments based on existing scripts. 

Throughout the investigation, we analyze its adaptability to various user interfaces, app architectures, and interaction patterns, ensuring accurate script generation and compatibility. The findings of this research contribute to the understanding of LLMs' capabilities in test automation. Ultimately, this research aims to enhance software testing practices, empowering app developers to achieve higher levels of software quality and development efficiency.

\end{abstract}

\vspace{1.5ex}
\begin{keywords}
\itshape Large Language Model, ChatGPT, Mobile App Testing, Test Generation, Test Migration
\end{keywords}

\section{Introduction}

With the rapid development and widespread adoption of mobile applications (apps), ensuring their quality and reliability has become increasingly vital \cite{yu2021layout}. Software testing has emerged as a dominant technique for quality assurance in the app development process. Test scripts, which serve as a set of predefined instructions and expected outcomes for conducting tests, play a critical role in the software testing process \cite{li2022mobile}. App developers create these scripts to automate the execution of test scripts and ensure consistent and repeatable results \cite{li2017atom}. However, generating and maintaining these test scripts poses several challenges in the dynamic and rapidly evolving mobile app landscape \cite{gao2015sitar}.

One of the significant challenges is the compatibility of test scripts with evolving app versions \cite{yu2021layout} \cite{gao2015sitar}. As developers release updates to their mobile apps, the underlying codebase and user interface may undergo significant changes. This often requires corresponding modifications to the test scripts to ensure their continued effectiveness during test generation. Failure to update the scripts may result in false positives or false negatives, leading to inaccurate test results and potentially overlooking critical defects.

Furthermore, the vast array of mobile devices, operating system versions, and screen sizes adds complexity to test script generation and maintenance \cite{yu2021layout} \cite{qin2019testmig} \cite{lin2022gui} \cite{liu2022test}. Mobile apps must function seamlessly across various platforms and form factors, requiring comprehensive testing across different device configurations. Generating and maintaining test scripts that cover this diverse landscape requires constant monitoring and updates to account for new devices, operating system versions, and resolution variations.

Test script generation and migration across different platforms, such as Android and iOS, presents significant challenges for existing test migration techniques \cite{yu2021layout}. Existing techniques, which capture user interactions as test scripts, rely heavily on the underlying platform structure and capabilities \cite{lin2022gui} \cite{liu2022test}. As a result, the generation and migration process introduces compatibility issues, rendering the recorded test scripts unusable or requiring extensive modifications. This is the consequence of programming language and framework gaps \cite{mariani2021semantic}, varying user interface guidelines and design patterns, and platform-specific features and functionalities of testing frameworks.

Another popular sub-topic of test script generation and migration is the migration among different apps sharing similar functionalities \cite{behrang2018test} \cite{behrang2019test}. While these apps may have similar features and user interactions, their underlying codebases and implementations can differ significantly. This variance poses difficulties for test scripts recorded in one app to be effectively applied to another. One challenge lies in the differences in the app's architecture and design patterns \cite{mariani2021semantic}. Even if two apps offer similar functionalities, their internal structures and coding practices may vary \cite{lin2019test}. This means that the recorded test scripts, which capture interactions specific to one app's architecture, may not align with the design patterns and structure of the target app. Consequently, extensive modifications are often necessary to ensure the test scripts accurately reflect the interactions and behaviors of the new app. Another challenge arises from the variations in the GUI design \cite{gao2015sitar}. While two apps may provide similar functionalities, their GUI layouts, navigation flows, and visual elements can differ significantly. Test scripts generated for one app may rely on specific GUI elements or sequences of interactions that are not present or operate differently in the target app. As a result, the test scripts must be carefully adjusted to accommodate these differences and accurately reproduce user interactions \cite{yu2021layout}.

These challenges have been deeply investigated and the solutions are improved, including some advanced learning based approaches, \eg deep learning, reinforcement learning \etc However, the effectiveness of existing approaches still exist bottlenecks. One of the most recent and promising AI technologies in this field is the large language models (LLM), including the Generative Pre-trained Transformer (GPT) developed by OpenAI\footnote{\url{https://chat.openai.com}}. LLMs have shown significant potential in various applications in software engineering \cite{ozkaya2023application}, including specification generation, developer assistance, unit test case generation, language translation, code summarization, and question answering, \etc

By leveraging the power of LLMs, it enables the automation of the test script generation and migration process. LLMs can understand and generate human-like text, making it capable of creating test scripts based on given scenarios. Furthermore, it is supposed to migrate these scripts, interact with the apps under test (AUT) as a human tester would, and provide detailed reports on the test results.

In this paper, we conduct an investigation on the capabilities of LLM in the mobile app test script generation and migration task, which is an extensively studied topic in the software testing community. Specifically, we use the \gpt as a representative of all LLMs. We conduct the investigation on three directions of test script generation and migration, including the scenario-based test generation, the cross-platform test migration, and the cross-app test migration. The scenario-based test generation focuses on the generation of test scripts based on the descriptions on a scenario of the AUT in natural language form. The cross-platform test migration focuses on the test script migration among different devices, operating systems, and platforms, including Android and iOS. The cross-app test migration focuses on the test script migration among apps sharing similar functionalities, which is more complex due to the differences among different apps.

Overall, the results of our investigation show that LLMs have a strong capability in the test script generation and migration tasks of different complexities. The most noteworthy advantage is that LLMs can well understand the business logics of the AUT, and can real-time adjust its generation and migration process based on the states of the AUT. However, we still find some limitations of LLMs, including the context memory, API usage randomness, human effort requirement, \etc Therefore, we propose the challenges and opportunities of applying LLMs in mobile app test script generation and migration, as a reference for further academic research.

In this paper, we propose the following contributions:

\begin{itemize}
	\item This is the first work that look into the application of LLMs in the mobile app test script generation and migration tasks, and provide some insightful findings.
	\item We conduct a thorough investigation on the capabilities of LLMs in the mobile app test script generation and migration with different configurations.
	\item We provide valuable future directions on applying LLMs in mobile app test script generation and migration, including the challenges and opportunities.
\end{itemize}

\section{Background}

\subsection{Test Generation and Migration} 

The surge in mobile app usage has highlighted the critical need for rigorous testing processes to ensure app functionality, performance, and user experience across a vast array of devices, operating systems, and network conditions. Manual testing alone is no longer sufficient, given the time-consuming nature of the task and the inherent limitations in replicating real-world scenarios. To tackle these challenges, researchers and practitioners have turned to automated approaches for mobile app testing, particularly in the generation of test scripts \cite{anbunathan2014event} \cite{zeng2016automated} \cite{li2017droidbot} \cite{wang2020combodroid} \cite{choi2022scriptpainter} \cite{tanno2015test} \cite{dallmeier2014webmate} \cite{iyama2018automatically}. The concept of mobile app test script generation involves the systematic generation of test cases or scripts that mimic user interactions with the app. These scripts aim to cover various use cases, edge cases, and device configurations to uncover potential bugs or issues in the application.

One of the leading frameworks in mobile app test script generation is Appium\footnote{\url{https://appium.io/}}. Appium is an open-source, cross-platform automation tool that enables the creation and execution of mobile app tests on different operating systems, including Android and iOS. Other automated frameworks include Calabash\footnote{\url{https://github.com/calabash/calabash-android/}}, Espresso\footnote{\url{https://developer.android.com/training/testing/espresso/}}, UI Automator\footnote{\url{https://developer.android.com/training/testing/ui-automator/}}, Selendroid\footnote{\url{http://selendroid.io/}}, 
XCUITest\footnote{\url{https://developer.apple.com/documentation/xctest/}}, \etc

Such frameworks, including Appium, leverage standard automation APIs, such as WebDriver, to facilitate seamless interaction with mobile apps, regardless of their platform or programming language. Appium's popularity stems from its versatility and compatibility with different development frameworks, such as Native, Hybrid, and Web applications. By employing Appium or similar frameworks, testers can automate repetitive and time-consuming tasks, expedite test execution, and increase test coverage across a wide range of mobile devices. Furthermore, these frameworks offer features like gesture recognition, image recognition, and synchronization capabilities, empowering testers to simulate complex user interactions and validate app behavior under diverse scenarios.

Mobile app test script migration \cite{qin2019testmig} \cite{lin2022gui} \cite{lin2019test} \cite{mesbah2011invariant} \cite{rau2018poster} \cite{rau2018transferring}, a process of transitioning test scripts from one framework or platform to another, has also emerged as a crucial area of research and development. Test scripts, which are sets of instructions or code that automate the execution of tests, are a crucial component of the testing process. However, as mobile app development frameworks, platforms, and technologies evolve rapidly, the need to migrate test scripts from one framework or platform to another has gained prominence. Mobile app test script migration involves adapting, translating, or converting test scripts written for one framework or platform to work effectively on another. This process is essential when organizations decide to switch development frameworks, upgrade app development platforms, or expand app compatibility across various devices.

The migration of test scripts presents numerous challenges to developers and testers. Differences in programming languages, syntax, APIs, test framework capabilities, and architectural differences between the source and target frameworks can pose significant obstacles. Ensuring the accuracy and fidelity of the migrated test scripts, maintaining test coverage, handling compatibility issues, managing data dependencies, and dealing with platform-specific nuances are additional challenges that need to be addressed. Moreover, the time and effort required for script migration can impact project timelines and budgets. To overcome the challenges associated with mobile app test script migration, researchers and practitioners have proposed various solutions and techniques. These include automated script migration tools, code analysis, and transformation techniques, script adaptation approaches, framework-independent testing methodologies, and strategies for effective script refactoring. Leveraging these solutions can streamline the migration process, reduce manual effort, enhance script accuracy, and improve overall testing efficiency.

\subsection{Large Language Model}

Large language models, such as GPT-x, have gained significant attention in the field of natural language processing (NLP) due to their ability to generate coherent and contextually relevant text. These models are trained on vast amounts of text data and can understand and generate human-like responses to a wide range of prompts. LLMs are built using deep learning techniques, specifically a type of neural network called a transformer. Transformers are designed to handle sequential data, such as text, and have been highly successful in various NLP tasks. The core component of an LLM is the transformer network. It consists of multiple layers of self-attention mechanisms and feed-forward neural networks. The self-attention mechanism allows the model to weigh the importance of different words or tokens in a sequence when generating responses. Text input for LLMs is typically tokenized into smaller units, such as words or subwords. Each token is associated with a unique numerical representation called an embedding. Tokenization allows the model to process text efficiently and capture the relationships between different tokens. When the model is deployed for real-world use, it goes through an inference phase. During inference, the model takes a sequence of tokens as input and generates a probability distribution over the possible next tokens. The token with the highest probability is selected as the model's output.

GPT-x, as a group of representative LLMs, is a massive model with billions of parameters, which makes it one of the largest language models created to date. The large number of parameters enables the model to capture intricate language patterns and generate detailed and contextually relevant responses. GPT-3 is built upon the transformer architecture, which consists of multiple layers of self-attention mechanisms and feed-forward neural networks. Self-attention allows the model to focus on different parts of the input sequence during training and inference, capturing long-range dependencies and contextual relationships. GPT-x is pre-trained on a vast amount of text data from the internet, including books, articles, websites, and other publicly available sources. After pre-training, GPT-x can be further fine-tuned on specific tasks or domains. Fine-tuning involves training the model on a more focused dataset with labeled examples or a reward-based setup. Fine-tuning helps customize the model for specific applications and improves its performance on targeted tasks.

\section{Study Design}

\subsection{Research Questions}

In this paper, we research on three research questions for mobile app test script generation and migration.

\begin{itemize}
	\item \textbf{RQ1 (Scenario-based Test Script Generation)}: how well do LLMs perform in generating test scripts based on the natural language descriptions on specific app scenarios?
	\item \textbf{RQ2 (Cross-platform Test Script Migration)}: how well do LLMs perform in the test script migration across different platforms for the same app (\eg Android, iOS)?
	\item \textbf{RQ3 (Cross-app Test Script Migration)}: how well do LLMs perform in migrating test scripts among different apps sharing similar functionalities?
\end{itemize}

These three research questions are progressive. The first research question is to simulate human testers who face the testing requirements documents to develop test scripts. The second research question is to simulate human testers when they need to migrate their existing test scripts of their apps among different versions on diverse platforms. The third research question is to simulate human testers when they need to migrate test scripts among different apps sharing similar functionalities, like the banking apps, SMS apps, \etc 
The first research question remains hard for automated techniques because the business logics of AUT are hard to analyze. The second and third research question has been solved to some extent, by analyzing the GUI layout files or GUI images. But the limitations are obvious, which we will discuss in \secref{sec:co}.

\subsection{LLM Setup} 

In our investigation, we use the API provided by OpenAI with the \texttt{gpt-3.5-turbo} model as a representative of LLMs. For the same research question, we generate different prompts and iteratively modify the prompts for better answers. Some prompts are repeated to see whether \gpt will generate responses with different scripts to eliminate the effect of randomness.

\subsection{Experiment Subjects}

In the context of our exploration into LLM for test script generation and migration, we have chosen a diverse array of widely used mobile apps as subjects for our experiment. These have been carefully selected based on their popularity, usability, and the varying complexity of their features. Our selection includes three email apps and three travel apps. For the email apps, we choose ``Outlook'', ``QQ Mail'', and ``NetEase Mail''. On the other hand, for travel apps, we pick ``Fliggy'', ``Ctrip'', and ``Mafengwo''. These chosen mobile apps are readily available and can be easily downloaded from various mobile app stores, such as Google Play Store and HUAWEI AppGallery.

\section{Result Analysis}

\subsection{RQ1: Scenario-based Test Script Generation}

Test scripts are usually developed for certain scenarios in practice. To evaluate the capability of LLMs in scenario-based test script generation, we have implemented an intuitive procedure for end-to-end script generation that closely mirrors the real-life situations encountered in the development of test scripts. First, we manually acquire comprehensive information of the whole testing process, including but not limited to essential configuration data (package name of AUT, name of testing device, \etc) and operational process details (identifiers of target elements, description of operations, \etc). Then, we provide \gpt with all information at once and instruct it to generate an corresponding test script.

The way we prompt the LLM is demonstrated in the following template:

\begin{GPT}{General Prompt}
\begin{itemize}
    \item Here are the initial values: appium:deviceName=\{\}, appium:appPackage=\{\}, appium:appActivity=\{\}, appium:noReset=\{\}, appium:fullReset=\{\}
    \item Page1: Click the ``\{\}'' button (ID: ``\{\}").
    \item Page2: Pass ``\{\}'' to the ``\{\}'' text box (ID: ``\{\}''). Then, pass ``\{\}'' to the ``\{\}'' text box (ID: ``\{\}''). Next, pass ``\{\}'' to the ``\{\}'' text box (XPath: ``\{\}''). Finally, click the ``\{\}'' button (ID: ``\{\}'').
    \item ...
    \item Use the above information to generate a Python test script executable on the device. Ensure to set a wait time where loading is required.
\end{itemize}
\end{GPT}

We conduct nine experiments of varying complexity, ranging from simple scenarios like ``login'' to more intricate scenarios like ``sending email'' and ``searching and booking flight ticket'', each performed on three distinct apps. Evaluation of the generated scripts is based on grammatical accuracy, semantic correctness, and practical applicability.

\gpt succeeds in creating corresponding Appium scripts by our instructions. All scripts are both grammatically and syntactically correct, and their alignment with the predefined test operation process is manually confirmed. However, there are also some technical problems during the direct execution of the scripts. These issues include the use of incompatible Appium APIs, random pop-ups interrupting script execution, and improper handling of input box focus during input operations. To be specific, \gpt uses three different APIs to locate elements across the generated scripts, which targets at the same operation:

\begin{enumerate}
    \item wait.until(EC.presence\_of\_element\_located((By.ID, ``id'')))
    \item driver.find\_element(By.ID, ``id'')
    \item driver.find\_element\_by\_id(``id'')
\end{enumerate}

and the third of which is deprecated in the latest version of Appium. Second, unexpected pop-ups arising during test script execution can cause failures, which come from an inherent vulnerability of test scripts that requires additional handling strategies. Third, input operations occasionally require initial focus on the target input box to ensure successful new input, such as transitioning focus from the username to the password input box. The scripts generated by \gpt are relatively random in handling this detail, failing to consistently meet this requirement.

Our preliminary experiments affirm that advanced LLMs can manage test script generation in scenarios of different complexity levels if provided with enough information. While some detected issues require manual adjustments, they are mainly technical and unrelated to the business logic under test.

To avoid the substantial manual effort to devise test procedures, we apply a dialogue-based approach to interactively guide LLM in AUT exploration and test script generation for specific scenarios, exploiting the abilities of contextual comprehension and robust logical reasoning from LLM. This approach significantly reduces manual intervention and enhances script generation efficiency, even with minimal supervision from non-specialists in software testing.

The dialogue is partitioned into three phases: initiation, exploration, and summarization. In the initiation phase, LLM is provided with the fundamental test context, including basic information about AUT, the specific testing scenario, and all the tasks for each iteration in the exploration phase. Task-1 asks LLM to determine whether the target test is complete - if so, the exploration phase concludes, otherwise, task-2 is initiated. Task-2 involves analyzing the provided XML elements and identifying the necessary test operation for this iteration, characterized by ``element-xpath'', ``operation-type'', and ``operation-text'' as keys in JSON format. In the exploration phase, LLM is required to analyze necessary XML elements present on the current page in each turn. It is then called upon to make consequential operational decisions for the successive steps in the test script. During this iterative and self-regulated process, the LLM correlates the provided elements with prior knowledge, identifies critical actions, and decides on the most effective paths to proceed. It adjusts its choices based on the outcomes, gradually optimizing the flow of actions, until it is assured that the testing requirements for the targeted function have been fully met. In the summarization phase, LLM is instructed to synthesize the entire exploration process, generating a comprehensive, detailed, and meticulously crafted test script that effectively satisfies the testing scenario as the final output.

The way we prompt the LLM in each phase is demonstrated in the following templates:

\begin{GPT}{Initiation Phase Prompt}
\begin{itemize}
    \item You are a software testing engineer.
    \item You are asked to test function ``\{\}''\space in app ``\{\}''.
    \item You will be provided with necessary XML structure of the current page each turn.
    \item You should perform the following tasks each turn:
    \begin{itemize}
        \item \textless TASK-1\textgreater\space Check whether the function has been tested. If true, summarize all the actions you have done and say ``DONE''. Otherwise, perform TASK-2.
        \item \textless TASK-2\textgreater\space Analyze the provided XML structure of the current page. If an appropriate element for operation can not be found, try drag operations. Describe what to do in JSON format with the following keys: ``element-xpath'', ``operation-type'', ``operation-text''.
    \end{itemize}
    \item Repeat what you are going to do and get ready.
\end{itemize}
\end{GPT}

\begin{GPT}{Exploration Phase Prompt}
\begin{itemize}
    \item Previous \textless click/input/drag\textgreater\space operation finished.
    \item Now we are in a new page. (Optional)
    \item The page remains unchanged. (Optional)
    \item \textless XML-element-1\textgreater
    \item \textless XML-element-2\textgreater
    \item ...
\end{itemize}
\end{GPT}

\begin{GPT}{Summarization Phase Prompt}
\begin{itemize}
    \item Generate Appium test script for the testing process.
\end{itemize}
\end{GPT}

To assess the effectiveness and efficiency of this novel approach, we conduct two separate experiments on the ``NetEase Mail'' app. Our focus is on two scenarios, ``login'' and ``adding email account'', with each subject to twice testing.

In the ``login'' scenario tests, \gpt consistently demonstrate a successful and efficient dialogic process, typically completed in seven to eight rounds. It can generate Appium test scripts similar to outputs in our preliminary experiments. This emphasizes \gpt's ability to create test scripts for specific functionalities by guiding with prompts. Furthermore, the model displays impressive competency in understanding and processing semantic information in XML element attributes such as \emph{context}, \emph{id} and \emph{hint}. Notably, \gpt demonstrates the capacity for reasoning and self-correction, identifying and correcting a non-responsive ``Login'' button issue arousing from an unchecked ``Agree to the Terms of Service'' box in the first test, and correctly navigating the process in the second test from the outset.

However, the ``adding email account'' scenario presents more of a challenge to \gpt. While it managed to navigate through the testing process successfully, \gpt fails to adhere to the established template when it should have generated the test script. The problem arises when \gpt, having completed the ``adding email account'' test, continues analyzing elements on the following page, making unnecessary action suggestions. A review of the context suggests that \gpt loses track of the prior context and fails to remember the original task, which leads to it being unable to self-terminate the testing process. The issue is likely due to the latest page containing an excessive number of clickable elements, making the whole context overwhelming \gpt's memory limit. When manually directed to the summarization phase, \gpt is only able to generate a script that lacked the initial test click in the first test. And in the second test, it fails to generate a relevant script at all.

\subsection{RQ2: Cross-platform Test Script Migration}

In some cases, the process of manual cross-platform test script migration necessitates an in-depth exploration of the differences that may arise within the target scenario across various brands or versions of a given app. This, in turn, requires a sequential examination and subsequent rectification of the script code to accommodate these differences. From researchers' perspectives, the introduction of LLM into this process is basically approached with a strategy to mimic human cognitive patterns. The objective is to equip LLM with sufficient information to facilitate the inspection, modification, or generation of scripts compatible with a new brand or version.

To evaluate the effectiveness of this approach, we carried out three sets of experiments, testing the performance of \gpt in cross-platform test script migration in the context of a ``login'' scenario on three different mobile apps. During these experiments, we endeavor to strike a balance between minimal and sufficient data provision in the migration task. By considering both experience of test script developers and  responses received from interactions with \gpt, we conclude the necessary information that needed to be presented to the LLMs, including new device name, new device version or brand information, differential steps, identifiers of the elements involved in differential steps (ID or XPath), and old test script. This set of information is crucial in helping LLM understand the differences between different platforms and enabled it to locate and interact with the elements on the new platform interface. For the purpose of the experiment, we provided all this information to \gpt in a single prompt and asked it to generate a new executable test script. The way we prompt the LLM is demonstrated in the following template.

In the process of assessing the new test scripts generated by \gpt, we execute them on the designated testing devices. Overall, the execution is predominantly seamless. Nevertheless, specific instances are noted where the ``entering password" operation cannot be successfully completed. This observation corresponds to a problematic scenario identified during the RQ1 experiments, which is the incorrect management of the focus on the input box during input operations. Therefore, it is necessary to make manual corrections after generation.

\begin{GPT}{General Prompt}
\begin{itemize}
    \item You are a software testing engineer.
    \item You are asked to do test script migration for a new platform.
    \item The information you know is list as follows:
    \begin{itemize}
        \item New device name: \{\}
        \item New Android version: \{\}
        \item Different steps:
        \begin{itemize}
            \item Step-1: \{\}
            \item Step-2: \{\}
            \item ...
        \end{itemize}
        \item Old test script: \{\}
    \end{itemize}
    \item Please return the new test script.
\end{itemize}
\end{GPT}

This experiment, to some extent, demonstrated the reliability of the code generated by \gpt. It exhibits impressive precision in the adaptation of the older script, leveraging the information supplied by human testers. However, it also reveals the constraints of \gpt, namely its inherent inability to self-correct or repair the script. It operates solely on the basis of human-provided information, blindly altering the old script without any inbuilt capacity for problem resolution if not prompted. This means that in migration scenarios with  intricate variances and challenges, dependence on the \gpt-generated code alone might not provide comprehensive solutions. In such circumstances, manual modifications and corrections based on experience of skillful test script developers may be required to tackle specific issues and enhance the robustness of the script.

Our further experiments indicate that any intentional exclusion of the provided data hinders the capability of \gpt to generate a comprehensive new script. However, when the absent information is manually reinstated during the interaction with \gpt, it proves to be able to generate a complete and precise migrated test script. This discovery highlights the significance of providing adequate information for the successful utilization of LLMs in cross-platform test script migration. Furthermore, the finding confirms we have employed a minimal information set required for LLMs to accomplish the task effectively.

\subsection{RQ3: Cross-app Test Script Migration}

In the task of cross-app test script migration, the adopted methodology parallels that used in cross-platform test script migration. This involves employing LLM to aid in the transfer of test scripts, which vary in complexity, from one application to a corresponding application within the same category. The primary objective is to facilitate automated script migration across similar applications, thereby reducing manual labor. As for the construction of prompts, we concentrate on distinguishing and highlighting the difference in testing procedures and related elements between the original and the target application.

\begin{GPT}{General Prompt}
\begin{itemize}
    \item You are a software testing engineer.
    \item You are asked to do test script migration for an app sharing the same function.
    \item The information you know is list as follows:
    \begin{itemize}
        \item New app information:
        \begin{itemize}
            \item Package name: \{\}
            \item Main activity name: \{\}
        \end{itemize}
        \item Different steps:
        \begin{itemize}
            \item Step-1: \{\}
            \item Step-2: \{\}
            \item ...
        \end{itemize}
        \item Old test script: \{\}
    \end{itemize}
    \item Please return the new test script.
\end{itemize}
\end{GPT}

In the experiment, three different applications - ``Outlook'', ``QQ Mail'', and ``NetEase Mail'' - are utilized for the ``login'' and ``sending email'' scenarios. For the ``flight search and reservation'' scenario, we use ``Fliggy'', ``Ctrip'', and ``Mafengwo''. These applications encapsulate a range of varying complexity levels of test scenarios. For every distinct scenario, a specialized testing script is manually crafted for the corresponding application, then we use \gpt to transpose this script onto applications with analogous scenarios.

The migrated testing scripts are subsequently evaluated through their execution on the designated applications, coupled with manual verification of the business logic's correctness and thorough coverage of the target scenarios. Based on our experimental observations, \gpt demonstrates no significant deviation in its ability to test scenarios of varying complexity. Provided there is an ample supply of information, \gpt shows efficiency in accomplishing cross-app test script migration tasks.

However, a notable limitation of \gpt lies in its lack of intrinsic understanding specific to software testing, compared to specialized testing tools or experienced testers and developers. It does not possess the capacity to autonomously acquire or refresh data about the target applications, which creates a reliance on continuous prompting with accurate and comprehensive information and instructions. All the code incorporated in the scripts generated by \gpt is heavily dependent on the prompt provided by the test personnel. Similar scenarios in different apps share similar core business logic, but test procedures may vary greatly due to differentiated design philosophies and implementation approaches. Therefore, it requires significant modifications to the existing scripts for similar scenarios across diverse apps. The trial in cross-app test script migration can therefore be perceived as creating an entirely new test script. For LLMs, which rigidly rely on prompt for generation, the migration task essentially transforms into a few-shot generation task that uses an existing test script as an example. Interacting with \gpt involves a significant investment of time for testers, who need to identify and organize information suitable for prompting. This process can be time-consuming and potentially outpace the time required for manual testing.

\section{Challenges, Capabilities \& Opportunities}
\label{sec:co}

In this section, we conclude the challenges faced by LLMs in mobile app test script generation and migration, and then we conclude the capabilities according to our investigation. Finally, we propose the opportunities of utilizing LLMs to automatically generate and migrate mobile app test scripts based on natural language descriptions in the future.

\subsection{Challenges}

\textbf{Context Memory}. The first significant challenge when LLMs are required to generate and migrate test scripts for mobile apps is primarily due to their limitations in retaining the testing context. While LLMs possess an extraordinary ability to generate human-like text, their inherent weakness lies in their struggle to remember important details and maintain a coherent understanding of the task at hand. This deficiency poses a hurdle when it comes to creating accurate and contextually relevant test scripts, especially within the dynamic and complex environment of mobile app testing. Therefore, despite their immense potential, LLMs still need further refinement to overcome these challenges and ensure seamless and efficient test script generation and migration for mobile applications.

\textbf{API Usage Randomness}. The second challenge of the utilization of LLMs in generating and migrating test scripts for mobile apps is due to their inability to consistently produce directly executable scripts. One of the reasons behind this limitation is that LLMs might employ inappropriate APIs, resulting in scripts that cannot be executed as intended. Additionally, since LLMs are trained on vast amounts of data, they may inadvertently incorporate deprecated APIs, which further hampers their effectiveness in generating functional and up-to-date test scripts. As a result, careful manual intervention and verification are necessary to ensure that the generated scripts align with the current APIs and can be seamlessly executed within the mobile app testing environment.

\textbf{Human Effort Requirement}. The third challenge of LLM is their reliance on human involvement to render the scripts executable. Despite their impressive capabilities, LLMs often require human intervention to fine-tune the generated scripts, incorporating essential configurations and modifying widget identifiers to align with the specific mobile app's structure. The dynamic nature of mobile app testing demands a deep understanding of the app's intricacies and peculiarities, which LLMs may not fully grasp. Therefore, human expertise is indispensable in refining and adapting the generated scripts, ensuring their compatibility with the app's environment and maximizing their effectiveness in the testing process.

\textbf{Limited Supported Test Events}. The fourth challenge LLMs encounter in mobile app test script generation and migration due to their limited support for certain test events. While LLMs excel at generating scripts for common interactions like clicking and inputting data, they often fall short when it comes to handling more complex events such as scrolling or managing pop-up windows. These additional test events require specialized knowledge and a real-time understanding of the app's user interface and behavior, which LLMs may not possess. As a result, manual intervention or the integration of complementary tools becomes necessary to address these gaps and ensure comprehensive test coverage for mobile app testing. Overcoming these limitations will require further research and development to enhance the capabilities of LLMs in generating test scripts that encompass a wider range of test events.

\subsection{Capabilities}

\textbf{Business Logic Understanding}. One of the key advantages of LLMs in mobile app test script generation and migration lies in their ability to effectively comprehend the underlying business logic of the apps being tested. Unlike traditional automated approaches, LLMs can delve deeper into the intricacies of the app's functionality and generate test scripts that align with its specific requirements. Moreover, LLMs excel in generating meaningful input strings that facilitate comprehensive testing. By leveraging their vast knowledge and contextual understanding, LLMs can push the boundaries of testing, exploring different scenarios and uncovering potential issues that may go unnoticed by traditional approaches. This capability empowers testers to enhance the quality and effectiveness of their testing efforts, making LLMs a valuable asset in mobile app testing and migration processes.

\textbf{Multi-to-Multi Event Mapping}. LLMs offer distinct advantages in mobile app test script migration due to their ability to comprehend the test script as a holistic entity. This comprehensive understanding enables LLMs to achieve multiple-to-multiple test event mapping, a capability that surpasses the limitations of traditional approaches. While traditional methods typically facilitate one-to-one test event mapping, LLMs have the potential to handle complex scenarios where multiple test events interact with each other. By considering the interconnectedness of various test events, LLMs can facilitate more accurate and comprehensive test script migration, ensuring that the intricacies and dependencies within the test script are preserved. This capacity to handle multiple-to-multiple test event mapping contributes to the efficiency and accuracy of mobile app test script migration.

\textbf{Multi-level Granularity Prompting}. The advantages of LLMs in mobile app test script generation and migration are prominently manifested through their ability to leverage Multi-level Granularity Prompting. LLMs possess the capability to consider various levels of granularity, such as the script-as-a-whole level or the test event level, to enhance the quality and effectiveness of the generated test scripts. By combining these different levels of granularity, LLMs can capture the nuances and intricacies of the testing requirements more comprehensively. This approach enables LLMs to generate test scripts that align with the desired objectives, taking into account the overall script structure as well as the specific details of individual test events. This flexibility and adaptability in incorporating Multi-level Granularity Prompting contribute to the accuracy and relevance of the generated test scripts, making LLMs an invaluable tool in mobile app test script generation and migration.

\subsection{Opportunities} 

\textbf{GUI Information Understanding}. The further enhancement of LLMs can be achieved by enabling them to directly comprehend GUI images from a visual perspective, akin to a human tester. While LLMs possess remarkable text-based comprehension capabilities, their understanding of GUI elements is currently limited. By incorporating visual understanding and analysis into LLMs, they could interpret and extract valuable information from GUI images, recognizing UI components, layouts, and visual cues. This enhancement would bridge the gap between LLMs and human testers, facilitating a more comprehensive understanding of the app's visual context and enabling LLMs to generate test scripts that accurately reflect the GUI's visual aspects. Integrating visual comprehension into LLMs would unlock their full potential in mobile app testing, providing a more holistic and efficient approach to test script generation and migration.

\textbf{Multi-modal Information Fusion}. The advancement of LLMs can be further realized by incorporating the fusion of multi-modal information, encompassing natural language documents, semi-structured logs, source code, and image data. By integrating these diverse modalities, LLMs would gain a more comprehensive understanding of the application under scrutiny. Natural language documents would provide contextual information, while semi-structured logs would offer valuable insights into system behavior. Analyzing source code would enable LLMs to grasp the underlying technical intricacies, and image information would grant them visual comprehension of the GUI. By fusing these multi-modal inputs, LLMs would possess a more holistic view of the application, enhancing their ability to generate accurate and contextually relevant test scripts. This advancement would significantly elevate LLMs' effectiveness in mobile app testing, enabling them to leverage a wider range of information sources and produce more comprehensive and robust test scripts and migration strategies.

\textbf{Domain Knowledge}. To further enhance the LLMs, a possible effective approach would involve fine-tuning them with domain-specific data, including the invaluable domain knowledge derived from human testers. Open-source models like Llama\footnote{\url{https://ai.meta.com/llama/}} and StarChat\footnote{\url{https://huggingface.co/HuggingFaceH4/starchat-alpha}} have enables the fine-tuning based on pre-trained LLMs. By incorporating this specialized knowledge, LLMs can gain a deeper understanding of the intricacies and nuances specific to the targeted domain. Human testers possess valuable insights into the domain, accumulated through their experience and expertise. By leveraging their knowledge to fine-tune LLMs, the models can acquire a more accurate and domain-specific understanding of the testing requirements, ensuring the generated test scripts and migration strategies align closely with the intricacies of the domain. This collaborative approach, combining the power of LLMs with human expertise, has the potential to significantly enhance the effectiveness and efficiency of mobile app test script generation and migration, ultimately improving the quality and robustness of the testing process.

\section{Related Work}

In this section, we introduce the related work of this paper. The studies are organized as three parts: test script generation, test script migration, and LLM for software engineering tasks.

\subsection{Test Script Generation}

One group of approaches focus on the automated exploration of the apps under test, including random-based, model-based, and learning-based strategies.

One of the fundamental approaches is the random-based strategy, which incorporates various tools. Among them, Monkey stands out as the most prevalent tool \cite{google2022monkey}, demonstrating commendable performance in specific benchmark applications \cite{choudhary2015automated}. Monkey boasts remarkable efficiency in generating a substantial volume of test events. Nevertheless, it exhibits evident drawbacks. The absence of effective guidance leads to the generation of a significant proportion of non-effective or redundant test events, thereby compromising the efficacy of the testing process. Although restraining GUI states helps mitigate this issue \cite{daragh2021deep}, the exploration still lacks proper guidance.

Various strategies have been employed to enhance testing exploration, such as the model-based approach \cite{amalfitano2014mobiguitar} \cite{yu2015incremental}, which involves constructing dedicated models using static, dynamic, or combined app analyses \cite{mesbah2011invariant}. Notable among these strategies is Sapienz \cite{mao2016sapienz}, a tool developed by Mao \etal, which utilizes multi-objective search-based testing to automatically explore and optimize test sequences. Another prominent model-based tool is Stout \cite{su2017guided}, which employs a stochastic Finite State Machine model to depict the behavior of the application under test. However, due to the limitations of the modeling algorithms, the model-based approach fails to accurately and comprehensively capture the intricacies of the tested applications. Aimdroid \cite{gu2017aimdroid}, introduced by Gu \etal, offers a novel approach by managing activity exploration and minimizing unnecessary transitions through an activity-insulated multi-level strategy during testing. In a similar vein, DIG \cite{biagiola2019diversity} adopts a pre-selection technique, selecting the most promising test cases based on their diversity from previous tests.

The rapid advancements in deep learning and machine learning models have paved the way for the emergence of learning-based automated testing technologies. Expanding on this notion, White \etal \cite{white2019improving} present a machine learning-driven approach that enhances GUI testing by automatically identifying GUI widgets within screenshots. In a similar vein, Li \etal \cite{li2019humanoid} introduce Humanoid, a deep learning-based automated black-box Android app testing tool. Reinforcement learning stands out as a particularly suitable algorithmic approach for app exploration. Koroglu \etal \cite{koroglu2018qbe} propose a fully automated black-box testing methodology that employs Q-Learning to explore GUI actions. Lv \etal \cite{lv2022fastbot2} contribute to this field by introducing an accelerated automated model-based GUI testing technique, showcasing its effectiveness through experiments conducted on two prominent industrial apps. Q-testing, a powerful Android testing tool \cite{pan2020reinforcement}, leverages RL algorithms to train a neural network capable of categorizing different states at the functional scenario level. In a related study, Romdhana \etal \cite{romdhana2021deep} conduct a comparative analysis of various RL algorithms in the context of mobile app testing.

Despite the advancements made in the realm of automated testing, significant challenges persist that impede the attainment of superior performance in terms of effectiveness and efficiency. Current approaches continue to grapple with critical issues that hinder their overall efficacy. Moreover, it is worth noting that none of the existing approaches have successfully achieved platform-independent app testing, presenting a substantial hurdle for app developers who contend with an ever-expanding array of operating platforms.

Except for some frameworks for test script generation \cite{wang2012test} \cite{anand2012automated} \cite{xu2014automated}, \eg Appium, some studies starts from the test script / case generation from the test results. 

Anand \etal \cite{anbunathan2014event} systematically and automatically generated test scripts based on concolic testing. Zeng \etal \cite{zeng2016automated} develop a new approach that inherits the high applicability of Monkey while addressing its empirically observed limitations. Li \etal \cite{li2017droidbot} introduce a new lightweight tool to automate mobile test input generation. Li \etal \cite{li2022mobile} present a new method that generates mobile test scripts from natural language. Wang \etal \cite{wang2020combodroid} generate high-quality mobile test cases by concatenating a number of use cases. Choi \etal \cite{choi2022scriptpainter} conduct research on mobile test script generation by analyzing videos. Tanno \etal \cite{tanno2015test} propose an approach to automatically generate test scripts from design documents, which are artifacts of the design process, using model-based testing. Dallmeier \etal \cite{dallmeier2014webmate} propose a method for automatically generating test scripts directly from a test target. Iyama \etal \cite{iyama2018automatically} propose a design that makes improvement by generating automatic test scripts without requiring additional man-hours for prior preparation and tool operation. 

\subsection{Test Script Migration}

Test script migration encompasses various aspects, including the migration among apps with similar functionalities within the same platform, as well as the migration of software with identical or similar functionalities between different platforms.

In the realm of cross-platform compatibility, several studies have addressed the challenge of migrating test scripts between web applications. Mesbah \etal \cite{mesbah2011invariant} and Rau \etal \cite{rau2018poster} \cite{rau2018transferring} focus on this area by exploring methods for transferring test scripts across different web platforms. Qin \etal \cite{qin2019testmig} propose an innovative approach to migrate GUI tests from iOS to Android without relying on migrated code samples. Another noteworthy contribution by Lin \etal \cite{lin2022gui} introduced an automated tool that transfers GUI tests from web applications to their Android counterparts. In addition, CraftDroid \cite{lin2019test}, have successfully migrated complex test cases, highlighting the potential of test reuse techniques in this context.

Liu \etal \cite{liu2022test} have conducted research on semantic matching in the context of GUI testing. Mariani \etal \cite{mariani2021semantic} \cite{mariani2021evolutionary} perform an empirical study on the semantic matching of GUI events, aiming to develop test reuse approaches that can automatically migrate human-designed GUI tests among apps with similar functionalities. Behrang \etal \cite{behrang2018test} \cite{behrang2019test} have placed more emphasis on scenarios and the iterative enhancement of their technique called AppTestMigrator, leverageing GUI similarities to facilitate test migration. Other researchers have approached the migration problem from the perspective of automatically repairing or patching migrated test scripts \cite{li2017atom} \cite{gao2015sitar} \cite{de2017reuse} \cite{thummalapenta2013efficient}. These efforts aim to enhance the usability and effectiveness of migrated tests by addressing potential issues through automated repair or patching mechanisms.

\subsection{LLM for Software Engineering} 

Large Language Models have the potential to be utilized in various software engineering applications. Moreover, with the rapid development of prompt engineering \cite{gao2023pal} \cite{liu2021generated} \cite{wang2022self} \cite{wei2022chain}, LLMs are increasingly playing a significant role in a wide range of areas of software engineering tasks \cite{sridhara2023chatgpt} \cite{tian2023chatgpt} \cite{mandal2023large} \cite{xing2023prompt} \cite{ross2023programmer} \cite{white2023chatgpt}.

Code generation is a major direction. During the era of statistical language models, code generation has already gone through a decent development \cite{raychev2014code}. However, LLMs can generate code snippets for developers based on natural language descriptions of desired functionality \cite{vaithilingam2022expectation} \cite{le2023methodology} \cite{macneil2023experiences}, ultimately saving more time and significantly improving the efficiency of the software development process. LLMs is also proved to be able to finish documentation generation tasks by utilizing natural language descriptions of the software functionality \cite{ahmed2022few} \cite{su2023hotgpt}.

Software testing is another field that LLMs can assist with \cite{allamanis2017learning}. Chen \etal \cite{chen2023teaching} propose Self-Debugging, which teaches a large language model to debug its predicted program via few-shot demonstrations. Kang \etal \cite{kang2022large} introduce LIBRO, a framework that uses LLMs to automate test generation from bug reports, showing promising results in reproducing and suggesting bug reproducing tests, thus enhancing developer efficiency. Schafer \etal \cite{schafer2023adaptive} introduce TestPilot, an adaptive test generation technique that utilizes LLMs to automatically generate unit tests for software programs, achieving high statement coverage and exercising functionality from the package under test, without requiring additional training or few-shot learning on existing tests. Feng \etal \cite{feng2023prompting} present AdbGPT, a lightweight approach that leverages LLMs to automatically reproduce bugs from bug reports through prompt engineering, using few-shot learning and chain-of-thought reasoning, achieving a high bug replay rate and outperforming existing baselines, as demonstrated by evaluations and confirmed by a user study, thus enhancing developers' bug replay capabilities. Liu \etal \cite{liu2022fill} introduce QTypist, an approach that utilizes a pre-trained LLMs to intelligently generate semantic input text for automated GUI testing of mobile apps, and when integrated with automated GUI testing tools, it covers more app activities, pages, and helps reveal more bugs compared to the raw tool.

\section{Conclusion}

The investigation in this paper makes a contribution to the field of mobile app testing. Firstly, it explores the capabilities of LLMs, represented by \gpt, in the generation and migration of test scripts. This sheds light on the potential of leveraging LLMs to automate these tasks and reduce the manual effort involved. The investigation focuses on three specific directions: scenario-based test generation, cross-platform test migration, and cross-app test migration. By examining these areas, the investigation provides insights into the challenges and complexities associated with generating and migrating test scripts in different contexts.

The investigation also identifies limitations, including issues related to context memory, API usage randomness, and human effort requirements, and thus provides valuable insights for future research and development in the field of mobile app test script generation and migration. Overall, the investigation contributes to advancing the understanding of the capabilities, limitations, and opportunities associated with using LLMs like GPT in the context of mobile app testing. It lays the foundation for further academic research and opens up possibilities for leveraging AI technologies to improve the quality and reliability of mobile apps. 

\section*{Acknowledgment}

This work is supported partially by the National Natural Science Foundation of China (62141215, 62272220, 62372228), the Science, Technology and Innovation Commission of Shenzhen Municipality (CJGJZD20200617103001003), and the National Undergraduate Training Program for Innovation and Entrepreneurship (202010284073Z).

\bibliographystyle{IEEEtran}
\bibliography{main}

\end{document}